\documentclass[aps,pra,numerical,superscriptaddress,showpacs,
reprint
]{revtex4-1}
\usepackage{amsmath}
\usepackage{color}
\usepackage{graphicx}
\usepackage[english]{babel}
\usepackage[noload=abbr]{siunitx}

\begin{document}
\title{Charge-induced optical bistability in thermal Rydberg vapor}

\author{Daniel Weller}
\affiliation{5.~Physikalisches Institut and Center for Integrated Quantum Science and Technology, University of Stuttgart, Pfaffenwaldring 57, 70569 Stuttgart, Germany}
\author{Alban Urvoy}
\affiliation{5.~Physikalisches Institut and Center for Integrated Quantum Science and Technology, University of Stuttgart, Pfaffenwaldring 57, 70569 Stuttgart, Germany}
\author{Andy Rico}
\affiliation{5.~Physikalisches Institut and Center for Integrated Quantum Science and Technology, University of Stuttgart, Pfaffenwaldring 57, 70569 Stuttgart, Germany}
\author{Robert L\"ow}
\affiliation{5.~Physikalisches Institut and Center for Integrated Quantum Science and Technology, University of Stuttgart, Pfaffenwaldring 57, 70569 Stuttgart, Germany}
\author{Harald K\"ubler}
\email{h.kuebler@physik.uni-stuttgart.de}
\affiliation{5.~Physikalisches Institut and Center for Integrated Quantum Science and Technology, University of Stuttgart, Pfaffenwaldring 57, 70569 Stuttgart, Germany}

\date{20 September 2016}
  
\begin{abstract}
We investigate the phenomenon of optical bistability in a driven ensemble of Rydberg atoms.
By performing two experiments with thermal vapors of rubidium and cesium,
we are able to shed light on the underlying interaction mechanisms
causing such a non-linear behavior.
Due to the different properties of these two atomic species,
we conclude that the large polarizability of Rydberg states
in combination with electric fields of spontaneously ionized Rydberg atoms
is the relevant interaction mechanism.
In the case of rubidium,
we directly measure the electric field
in a bistable situation
via two-species spectroscopy.
In cesium,
we make use of the different sign of the polarizability for different $l$-states
and the possibility of applying electric fields.
Both these experiments allow us 
to rule out dipole-dipole interactions,
and support our hypothesis of a charge-induced bistability.
\end{abstract}

\maketitle

\section{Introduction}
The extraordinary properties of highly excited atoms have led 
to a renaissance in the research field of Rydberg atoms.
Especially the strong interaction between Rydberg atoms
in combination with ultracold gases 
paved the way for
a variety of applications and novel phenomena,
among which are
quantum gates~\cite{saffman2005analysis,jones2007fast},
quantum phase transitions~\cite{low2009universal,schauss2014dynamical}, 
optical non-linearities on the single photon level~\cite{maxwell2013storage,dudin2012strongly,
peyronel2012quantum,tiarks2014single,gorniaczyk2014single}, 
beyond two-body interactions~\cite{faoro2015borromean},
excitation transfer~\cite{gunter2013observing,barredo2015coherent}
aggregation of excitations~\cite{schempp2014full,malossi2014full,urvoy2015strongly}
and ultralong-range molecules
~\cite{bendkowsky2009observation}.
Although most of the recent experimental results originate from ultracold atoms,
it is sometimes advantageous to study Rydberg atoms in thermal vapor,
especially when large atom numbers
or high number densities are required
--
as is the case for electric field sensing~\cite{sedlacek2013atom},
aggregation~\cite{urvoy2015strongly}
or optical bistabilities (detailed description in Sec.~\ref{sec:results}). 
Specifically the latter has exclusively been studied in thermal vapors
\cite{carr2013nonequilibrium,marcuzzi2014universal,
vsibalic2015driven,de2016intrinsic,ding2016non}.
In these publications,
one interpretation suggests 
that the underlying mechanism for the bistability
is a dipole-dipole interaction between Rydberg states. 
Our experimental findings presented here however
support an alternative interpretation:
We argue that the large polarizability of Rydberg states
in the presence of ions
is at the heart of the optical bistability in thermal Rydberg vapor. 

In this paper, we describe the observation of optical bistability
in two separate experiments,
one with a thermal vapor of rubidium atoms
and
the other with cesium.
We show that the underlying 
interaction shift
is caused by charges in the medium, 
themselves being produced by spontaneous ionization of Rydberg atoms. 
In the former case,
we make use of the natural abundance
of the two distinct rubidium isotopes.
While $^{85}$Rb atoms are driven to bistability, 
a simultaneous spectroscopic measurement on the coexisting $^{87}$Rb isotope 
shows a large broadening and shift of the Rydberg resonance. 
This observation is consistent with the presence of ions in the medium, 
induced by the excitation of Rydberg atoms. 
In the case of $^{133}$Cs, 
we observe a strong sensitivity of the optical bistability features
to external ac electric fields, 
strongly indicating that charged particles in the medium are closely related to the phenomenon. 
By compiling the signs of the interaction shifts for various species and Rydberg states, 
we finally conclude that the observed optical bistability 
is only compatible with interactions between Rydberg atoms and charges,
and a dipole-dipole interaction between Rydberg atoms is ruled out as a possible scenario.

The paper is organized as follows.
In Sec.~\ref{sec:MuM}
we describe the two experimental setups and the implementation of the measurements performed.
Because we used two different elements,
first the experiment with two rubidium isotopes is detailed first,
before we introduce the setup involving cesium.
Section~\ref{sec:results} then contains our analysis and argumentation
in the same order,
followed by the conclusion in Sec~\ref{sec:conclusion}.

\section{Materials and Methods}
\label{sec:MuM}
\subsection{Rubidium}
\begin{figure}[htbp]
    \includegraphics[width=\linewidth]{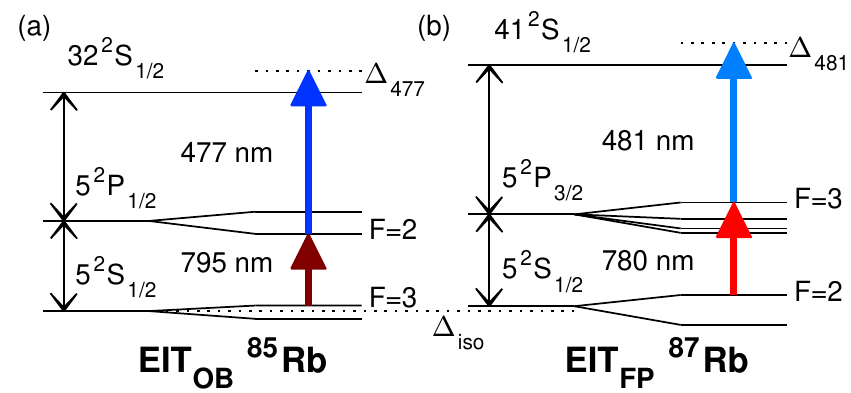}
    \caption{
      {Relevant transitions in the resonant two-photon excitation schemes for both isotopes.}
      \textbf{(a)} EIT$_\mathrm{OB}$ drives the optical bistability
      \textbf{(b)} EIT$_\mathrm{FP}$ acts as a probe for electric fields.}
    \label{fig:termschema}
\end{figure}

We drive an optical bistability in $^{85}$Rb 
similar to the work in Ref.~\cite{carr2013nonequilibrium}, 
using an EIT-like excitation scheme \cite{mohapatra2007coherent} -- 
depicted in Fig.~\ref{fig:termschema}(a) and referred to as EIT$_\mathrm{OB}$.
A frequency-stabilized \SI{795}{\nano\meter} laser probes the 
$^{85}$Rb $5$S$_{1/2}$, F$=3 \rightarrow 5$P$_{1/2}$, F$'=2$ transition on resonance
with a Rabi frequency of 
$\Omega_{795} = 2\pi\times\SI{37}{\mega\hertz}$, 
while a \SI{477}{\nano\meter} laser is scanned over the transition
\mbox{$5$P$_{1/2} \rightarrow 32$S$_{1/2}$}.
For the \SI{477}{\nano\meter} laser,
the maximal Rabi frequency is $2\pi\times\SI{25}{\mega\hertz}$,
limited by the available laser power.
The two laser beams are focused to a waist of \SI{40}{\micro\meter} 
and overlapped in a copropagating configuration.
This results in
a simple Lorentzian-like excitation spectrum, 
in contrast to the counterpropagating configuration
~\cite{shepherd1996wavelength,mohapatra2007coherent,urvoy2013optical}. 
Due to population shelving~\cite{thoumany2009optical}, 
the transmission signal of the \SI{795}{\nano\meter} laser is roughly proportional to the population of the Rydberg state.
The transmission of the \SI{795}{\nano\meter} laser is monitored 
as a function of the detuning $\Delta_{477}$,
recording both directions of the frequency scan
(from red to blue, and vice versa).

\begin{figure*}[htbp]
    \includegraphics[width=\textwidth]{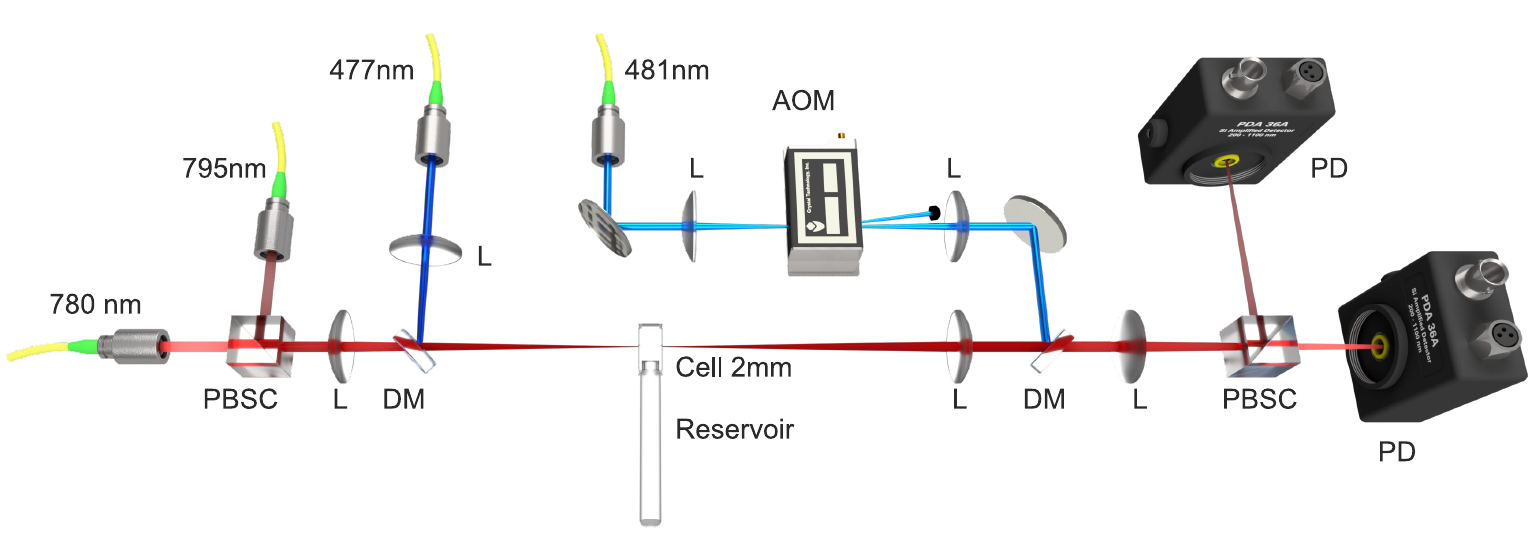}
    \caption{
      {Setup of the rubidium experiment.}
      Two pairs of lasers drive the two isotopes in the same volume
      using an EIT-like excitation scheme [see Fig.~\ref{fig:termschema}].
      One drives an optical bistability with the scheme EIT$_\mathrm{OB}$,
      while the other probes for local fields with the scheme EIT$_\mathrm{FP}$.}
    \label{fig:setup:rb}
\end{figure*} 
Figure~\ref{fig:setup:rb} shows the experimental setup.
A \SI{2}{\milli\meter} thick glass cell containing the rubidium vapor 
is placed at the focus of the beams.
The cell is a spectroscopy cuvette (quartz glass, Hellma Analytics) 
connected to a reservoir 
and filled with a droplet of naturally occurring rubidium 
(i.e., $78\%$ $^{85}$Rb and $22\%$ $^{87}$Rb)
under vacuum.
The temperature of the reservoir is stabilized to control the vapor density,
while the higher cell temperature (T$_\mathrm{res.}=$ \SIrange{80}{120}{\celsius}, T$_\mathrm{cell}=\SI{135}{\celsius}$) prevents unwanted condensation of the alkali.
By measuring the absorption in the D$_1$ line,
the vapor density is determined as
$\mathcal{N}_{85} = \SI{1.8e12}{\centi\meter^{-3}}$
and 
$\mathcal{N}_{87} = \SI{0.7e12}{\centi\meter^{-3}}$
for $^{85}$Rb and $^{87}$Rb, respectively.

In order to probe into the mechanisms 
and the cause of the optical bistability,
we simultaneously measure an additional EIT-like spectrum
\cite{mohapatra2007coherent}
on the less abundant isotope $^{87}$Rb,
hereafter named EIT$_\mathrm{FP}$ (\textit{field probe}). 
The advantage of using two different isotopes is that 
-- except for interspecies interactions --
the two schemes are completely decoupled
due to the different wavelengths of the transitions. 
Similar to the excitation scheme EIT$_\mathrm{OB}$ presented above, 
we use two lasers at \SI{780}{\nano\meter}
(probe, $\Omega_{780} = 2\pi\times\SI{66}{\mega\hertz}$) 
and \SI{481}{\nano\meter} 
(coupling, $\Omega_{481} = 2\pi\times\SI{10}{\mega\hertz}$)
 to drive the ladder scheme 
\mbox{$5$S$_{1/2}$, F$=2 \rightarrow 5$P$_{3/2}$, F$'=3 \rightarrow 41$S$_{1/2}$} 
of $^{87}$Rb 
[see Fig.~\ref{fig:termschema}(b)]. 
Again,
the \SI{780}{\nano\meter} laser is locked on resonance
and the \SI{481}{\nano\meter} laser has a variable detuning 
$\Delta_{481}$
with respect to the upper transition. 
These two additional laser beams are overlapped with the previous pair
and in the same arrangement as for EIT$_\mathrm{OB}$ up to the following: 
the \SI{780}{\nano\meter} and \SI{481}{\nano\meter} lasers are counter-propagating 
and the polarizations are perpendicular to those of EIT$_\mathrm{OB}$.
By independently scanning $\Delta_{481}$ and $\Delta_{477}$,
we obtain two-dimensional transmission spectra of the \SI{780}{\nano\meter} probe laser, 
an example of which is shown in Fig.~\ref{fig:2d}(a).
Since the choice of Rabi frequencies 
compromises between
a narrow linewidth and signal visibility,
the signal-to-noise ratio is improved 
by amplitude-modulating the \SI{481}{\nano\meter} laser with a frequency of \SI{30}{\kilo\hertz} 
and demodulating the transmission signal using a lock-in amplifier (Femto, LIA-MV-200).

\subsection{Cesium}
\begin{figure*}[htbp]
    \includegraphics[width=\textwidth]{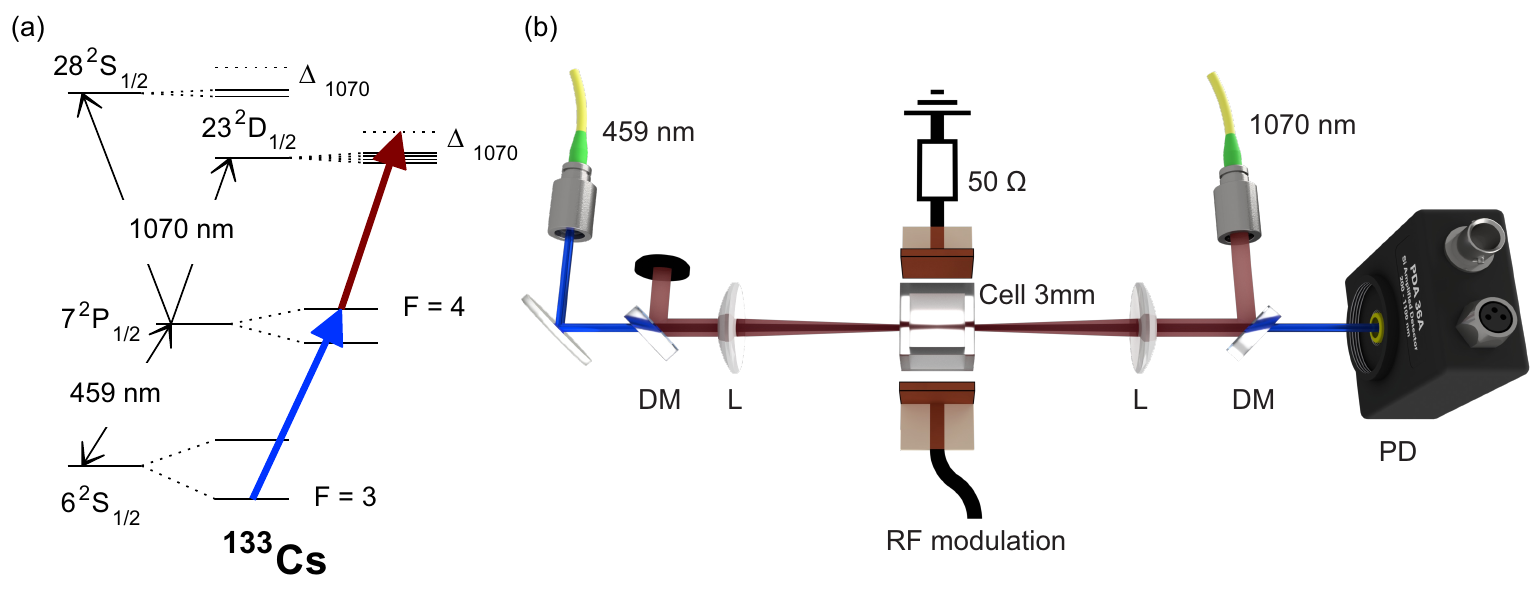}
    \caption{
      {Setup of the cesium experiment.}
          \textbf{(a)} Level scheme with the relevant cesium transitions.
          \textbf{(b)} Counter-propagating spectroscopy setup 
          with a \SI{3}{\milli\meter} thick cesium vapor sample.
          The cell is placed between two electrodes,
          where the modulation is applied on one port
          while the other is \SI{50}{\ohm} terminated.
          The laser polarization is parallel to the electric field between the plates.
      }
    \label{fig:setup:cs}
\end{figure*}

In the experiment based on $^{133}$Cs,
we address the atoms with the two-photon ladder scheme 
\mbox{$6$S$_{1/2}$, F$=3 \rightarrow 7$P$_{1/2}$, F$'=4 \rightarrow nl$} 
as depicted in Fig.~\ref{fig:setup:cs}(a).
The Rabi frequencies for the lower transition at \SI{459}{\nano\meter}
and the upper transition at \SI{1070}{\nano\meter}
are set to
$\Omega_{459} = 2\pi\times\SI{6}{\mega\hertz}$ 
and
$\Omega_{1070} = 2\pi\times\SI{146}{\mega\hertz}$,
respectively.
Similar to the rubidium experiment,
the transmission of the \SI{459}{\nano\meter} laser is measured directly on a photodiode
while the \SI{1070}{\nano\meter} detuning $\Delta_{1070}$ is varied.
The measurements are performed using counterpropagating lasers,
focused to \SI{50}{\micro\meter} in a \SI{3}{\milli\meter} vapor cell
($\mathcal{N}_\mathrm{Cs} = \SI{1.2e13}{\centi\meter^{-3}}$),
as sketched in Fig.~\ref{fig:setup:cs}(b).
In this inverted level scheme, 
where the wavelength of the lower transition is smaller than the one for the upper transition,
additional decay channels and transit time broadening result in enhanced absorption~\cite{urvoy2013optical}.
With the use of a fiber amplifier,
we can reach higher Rabi frequencies on the upper transition with cesium
than in our rubidium setup.
Furthermore, we can select various principal and two different azimuthal quantum numbers ($n=$ \numrange{20}{60}, \mbox{$l = S,D$}) for the Rydberg state,
thus examining the response of systems with different properties.
In the following,
we consider in particular the Rydberg states 23D$_{3/2}$ and 28S$_{1/2}$. 

The major additional feature available in this setup is the ability
to externally apply an electric field across the cell. 
The glass cell is placed between two electrodes 
that produce an electric field roughly parallel to the polarization of the laser beams.
A specially designed strip-line guides the applied microwave
(possible frequencies range from DC to several \SI{}{\giga\hertz})
to an electrode next to the spectroscopy cell.
The electrode on the other side is terminated with \SI{50}{\ohm} [Fig.~\ref{fig:setup:cs}(b)].
For our measurements,
we inject a sine wave with frequencies of 10 to \SI{500}{\mega\hertz},
yielding an oscillating electric field between the electrodes
with an amplitude of approximately \SI{3.2}{\volt\per\centi\meter}.

\section{Results and Discussion}
\label{sec:results}
\subsection{Rubidium}
\begin{figure}[htbp]
    \includegraphics[width=\linewidth]{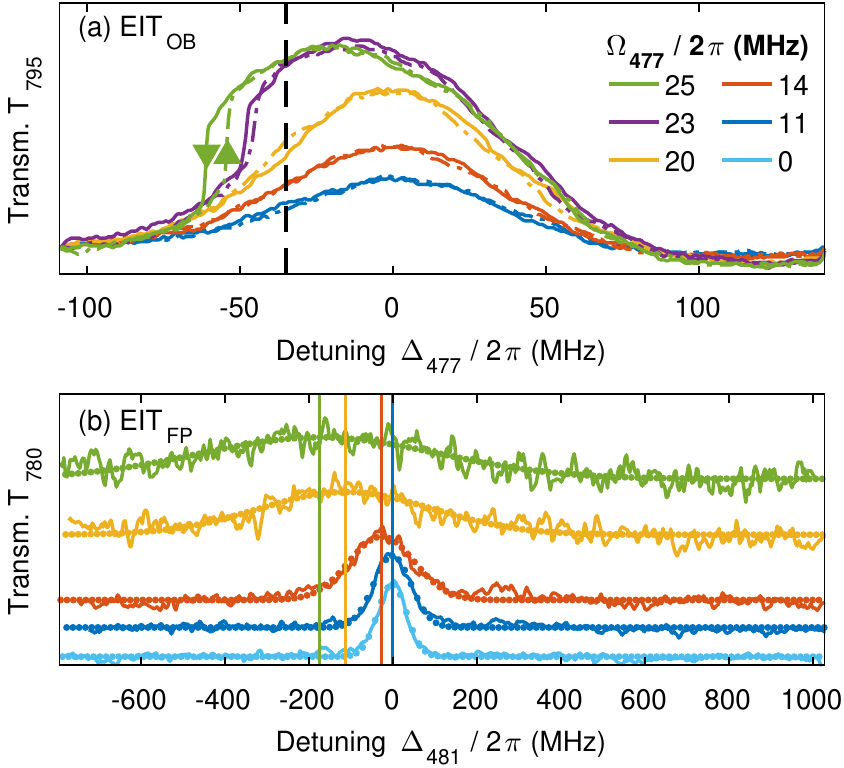}
    \caption{
      {Rubidium measurements.}
          \textbf{(a)} Deformation of the 795-laser (EIT$_\mathrm{OB}$)
          transmission peak towards optical bistability.
          Scans across the resonance from blue to red detuning (dashed)
          and reverse (solid line).
          \textbf{(b)} EIT$_\mathrm{FP}$ scans taken at a fixed detuning
          $\Delta_{477}/2\pi = \SI{-35}{\mega\hertz}$ [vertical dashed line in panel (a)].
          The feature is shown for different Rabi frequencies of the EIT$_\mathrm{OB}$ system
          (offset to discern the lines).
      }
    \label{fig:bistability}
\end{figure}

Figure~\ref{fig:bistability}(a) shows the typical traces of the EIT$_\mathrm{OB}$ system.
For increasing Rabi frequencies of the coupling laser,
the EIT$_\mathrm{OB}$ peak is shifted to the red and becomes more and more distorted,
until at sufficiently large intensity the system becomes bistable, 
as in Ref.~\cite{carr2013nonequilibrium}.
This phenomenon is the result of a competition between 
a non-linear energy shift due to an interaction effect
which is dependent on the Rydberg state population, 
and decay from the Rydberg state. 
On the one hand, when the frequency is scanned from the blue detuned side to the red,
there is a buildup of Rydberg population
that sustains the ability to excite Rydberg atoms even away from resonance.
When the detuning becomes too large,
the decay mechanisms prevail and the population breaks down. 
This results in a sudden change of the transmission level,
as can be observed in the solid lines in Fig.~\ref{fig:bistability}(a).
On the other hand, when the frequency is scanned to the blue (dashed lines), 
the Rydberg population stays low in the bistable region 
until the detuning becomes small enough to sufficiently excite atoms.
These atoms act like a seed for subsequent excitations.
When a certain threshold is reached,
a sudden increase in population is triggered
and the system is switched to the high population state.
Overall, a hysteresis in the transmission spectrum can be observed.

We make a complementary observation 
by probing the interatomic interactions in the excitation volume
using the additional Rydberg EIT scheme EIT$_\mathrm{FP}$.
For a fixed detuning $\Delta_{477}$,
we observe the signal in the EIT$_\mathrm{FP}$ scheme
as shown in Fig.~\ref{fig:bistability}(b). 
The traces here show a significantly growing shift
for increasing EIT$_\mathrm{OB}$ Rabi frequencies
and therefore a rising Rydberg population
in the bistable scheme EIT$_\mathrm{OB}$.
In Figs.~\ref{fig:bistability}(a) and \ref{fig:bistability}(b), 
besides a broadening of the lines,
the sign of the shifts in both schemes clearly indicates attractive interactions.

\subsubsection{Rydberg-Rydberg interactions}
\begin{figure}[htbp]
    \includegraphics[width=\linewidth]{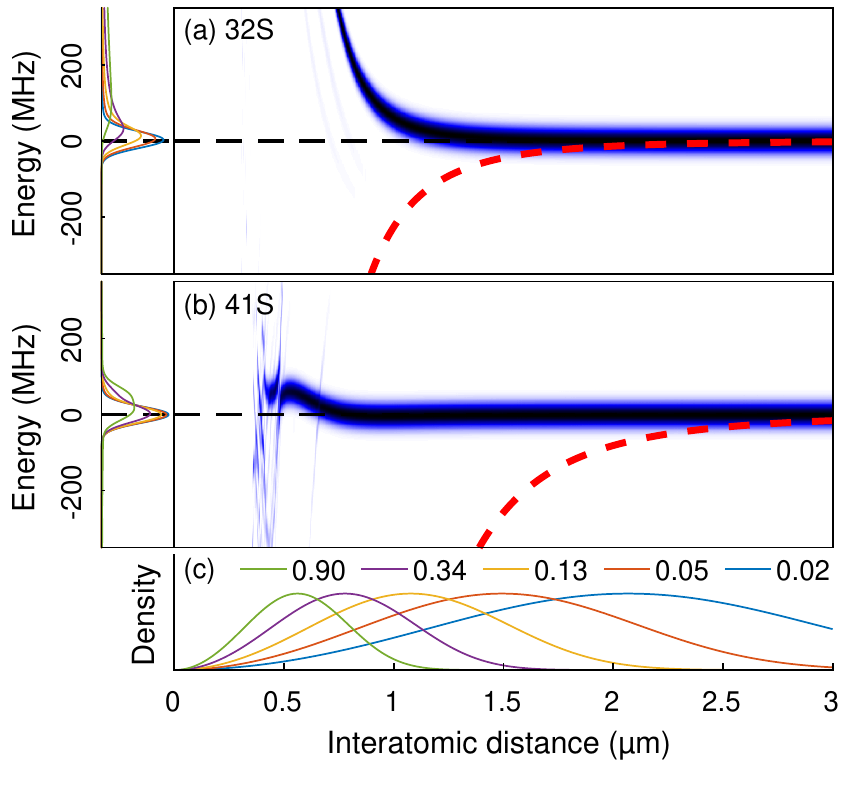}
    \caption{
      {Rubidium interaction potentials.}
          \textbf{(a)} Pair potentials for $32S$-$32S$.
          The color shading represents the projection on the unperturbed pair state.
          \textbf{(b)} Pair potentials for $41S$-$32S$. 
          The avoided crossings arise from the interaction with different pair-states
          (not shown here). 
          The $C_4/r^4$ potential is plotted as a red dashed line for the $32S$ state (a) 
          and the $41S$ (b).
          \textbf{(c)}
          Nearest-neighbor distribution for various densities
          (given in $\si{\micro\meter^{-3}}$) of Rydberg atoms, 
          corresponding to Rydberg fractions of \SIrange{1}{50}{\percent}. 
          The curves are normalized to the maximum density.
          \textbf{(left)} Projection of pair potentials weighted with distribution (c).
      }
    \label{fig:pp}
\end{figure}

Let us assume that Rydberg-Rydberg interactions 
(e.g.~dipole-dipole or van-der-Waals) are the underlying mechanism 
for the optical bistability.
Then, the interactions between $32S$ and $32S$ states on the one hand 
should explain the bistability itself,
and, on the other hand,
those between $41S$ and $32S$ are relevant
for the shifts measured with EIT$_\mathrm{FP}$.
Figures \ref{fig:pp}(a) and \ref{fig:pp}(b) show computed pair-potentials for the interactions 
of the $32S$ state
with the $32S$ and $41S$ states, respectively.
The potentials are calculated up to dipole-quadrupole terms
similar to those in Ref.~\cite{schwettmann2006cold}.
It is noteworthy that between rubidium $S$-states,
the van-der-Waals interaction potential is constantly repulsive,
i.e., leading to a shift towards blue wavelengths, contrary to what we observe.
In Fig.~\ref{fig:pp}(c),
the nearest-neighbor distance probability
(given by the Chandrasekhar distribution)
is plotted for various densities of Rydberg atoms. 
By weighting the pair-potential map with the nearest-neighbor distribution,
we determine the overall sign of the expected level shift in the experiment.
This gives a very rough estimate of the line shapes that one should expect from the measurements,
which are displayed on the left in Fig.~\ref{fig:pp}. 
Although this estimate only accounts for interactions between nearest neighbors,
it is clear that only blue shifts are to be expected with this mechanism.
The shift even vanishes for dilute vapors as given by our experimental parameters.
An estimate for the density of Rydberg atoms
is on the order of \SI{0.05}{\micro\meter^{-3}}. 

Apart from these direct interaction paths, 
it is possible that a significant part of the S-state population 
decays to neighboring $P$-states.
Therefore, also the interactions between states $32S$ and $31P$,
as well as between $41S$ and $31P$ are to be considered.
In a related experiment with ultracold atoms~\cite{goldschmidt2016anomalous},
it was shown that this configuration does not lead to a shift of the excitation spectrum. 
Only a broadening was observed,
consistent with the spatial integration of the dipole-dipole interaction potential in an isotropic medium. 
Also 
the $31P$-$41S$ dipole-dipole interaction is much weaker than the $31P$-$32S$ dipole-dipole interaction
because the wave function overlap is smaller. 
This comes in contradiction with the experimental observation in Fig.~\ref{fig:bistability}
that the interaction shifts on the EIT$_\mathrm{FP}$ scheme
are {larger} than those on the EIT$_\mathrm{OB}$ scheme.
It is therefore unlikely to explain the observed optical bistability
by means of Rydberg-Rydberg interactions.

\subsubsection{Ionization-induced electric fields}
As an alternative explanation for the observed red interaction shift, 
we suggest that the bistability is caused by charges created by the ionization of Rydberg atoms
~\cite{mohapatra2007coherent,vitrant1982rydberg,barredo2013electrical}. 
The interaction potential between a single charge
and a Rydberg atom (red dashed lines in Fig.~\ref{fig:pp}) 
in this picture arises from the dc Stark-shift of the Rydberg state and has the form $C_4/r^4$, 
where $C_4$ is proportional to the polarizability $\alpha$ of the Rydberg state.
Following the usual definition for the sign of the interaction,
a positive polarizability value yields a negative energy shift.
For rubidium $S$-states,
$\alpha$ is invariably positive~\cite{gallagher},
thus resulting in a red shift as observed.
The polarizabilities for the regarded states are computed as
\mbox{$\alpha_{32S} = \SI{2.2}{\mega\hertz\per(\volt\per\centi\meter)\squared}$} and
\mbox{$\alpha_{41S} = \SI{12.6}{\mega\hertz\per(\volt\per\centi\meter)\squared}$}, respectively.
Examining Fig.~\ref{fig:bistability} again given these numbers,
we find a good compatibility with both the sign and the magnitude of the observed shifts.
Since both isotopes are exposed to the same electric field distribution in the vapor,
the EIT$_\mathrm{FP}$ spectrum is shifted more,
according to the larger polarizability $\alpha_{41S}$.

\begin{figure}[htbp]
    \includegraphics[width=\linewidth]{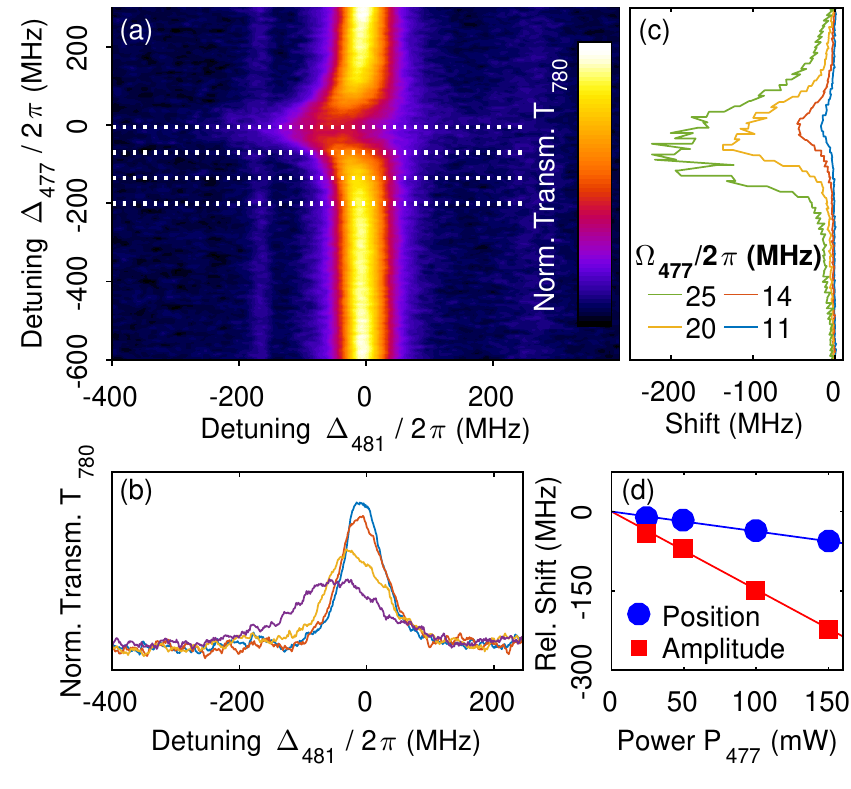}
    \caption{
      {Rubidium measurements.}
      \textbf{(a)}
      Example of a two-dimensional map showing the EIT$_\mathrm{FP}$ traces 
      against both detunings for $\Omega_{477}/2\pi = \SI{14}{\mega\hertz}$.
      \textbf{(b)}
      Cuts along the dashed lines in panel (a).
      \textbf{(c)}
      Dependence of the fitted maximum in every row in panel (a)
      for different Rabi frequencies $\Omega_{477}$
      \textbf{(d)}
      Center frequency and amplitude of the curves in panel (c), 
      determined via a Gaussian fit (uncertainty within marker size).  
      The center frequency (blue dots) represents the shift in the EIT$_\mathrm{OB}$ scheme, 
      while the amplitude (red squares) illustrates the shift in the EIT$_\mathrm{FP}$ scheme. 
      The offsets are chosen such that the linear fits intersect at zero.
      }
    \label{fig:2d}
\end{figure}

To further substantiate our hypothesis,
EIT$_\mathrm{FP}$ traces are systematically taken for a set of detunings $\Delta_{477}$ 
and for various Rabi frequencies $\Omega_{477}$ in the EIT$_\mathrm{OB}$ scheme.
The resulting data are displayed in Fig.~\ref{fig:2d}.
For an exemplary Rabi frequency of $\Omega_{477}=2\pi\times\SI{14}{\mega\hertz}$,
the resulting density plot of the transmission is shown in Fig.~\ref{fig:2d}(a), 
along with a sample of EIT$_\mathrm{FP}$ traces in (b). 
The behavior is very similar to what is shown in Fig.~\ref{fig:bistability}(b) 
for various Rabi frequencies $\Omega_{477}$, 
emphasizing the dependence of the EIT$_\mathrm{FP}$ signal on the excitation probability of $32S$ Rydberg atoms. 
We can gain further insight on the relation between the two systems by evaluating 
the shift in the EIT$_\mathrm{FP}$ system
relative to the unperturbed line. 
For a set of Rabi frequencies $\Omega_{477}$,
this shift 
(determined by the center frequency of a Gaussian fit to the EIT$_\mathrm{FP}$ signal)
is depicted in Fig.~\ref{fig:2d}(c)
against the EIT$_\mathrm{OB}$ detuning $\Delta_{477}$.
There is a clear evolution from the unshifted EIT$_\mathrm{FP}$ signal
in the off-resonant region 
to a maximally shifted signal,
where the EIT$_\mathrm{OB}$ system is close to resonance.
Noticeably,
the shift trajectories in the EIT$_\mathrm{FP}$ spectra 
show the same characteristic deformation and asymmetry towards red detuned frequencies  
as the transmission curves of EIT$_\mathrm{OB}$ in Fig.~\ref{fig:bistability}(a).
To conclude the analysis of Fig.~\ref{fig:2d},
both the amplitude and the center frequency (Gaussian fit to each curve)
of the shift trajectories from panel (c)
are plotted versus the Rabi frequency $\Omega_{477}$ [Fig.~\ref{fig:2d}(d)].
The amplitude here directly reflects the shift in the EIT$_\mathrm{FP}$ system, 
while the position is an indirect measure for the shift in the EIT$_\mathrm{OB}$ system.
The ratio between the slopes of the two linear fits amounts to 4:1. 
A qualitatively similar observation was made in Fig.~\ref{fig:bistability}
and again roughly reflects the ratio of polarizabilities between the states $41S$ and $32S$
which is 5.7:1. 

\subsubsection{Quantitative analysis}
Let us now first estimate the ion density present in the cell,
and then extrapolate the necessary cross section for the ionization process,
to allow a quantitative comparison to previous experiments.
In the scope of ionizing collisions,
the heavy particles remain quasi stationary
while the much lighter free electrons gain significantly more velocity. 
Hence, the electrons leave the volume in a much faster timescale compared to the heavier ions.
In a very simplified picture neglecting the free electrons and only considering the ions, 
the electric field distribution in the medium is described 
by the so-called Holtsmark distribution~\cite{holtsmark1919verbreiterung}.
The Holtsmark probability distribution function for the electric field
in this case is given by
\begin{equation}
  \mathcal{P}\left(E\right) = H\left(E/Q_H\right)/Q_H ,
\end{equation}
where the normal field is given by the expression
\begin{equation}
  Q_H = 
  \left(\frac{4}{15}\right)^{2/3} \left(\frac{e}{2\epsilon_0}\right)
  \left(\mathcal{N_\mathrm{ion}}\right)^{2/3} ,
\end{equation}
and
\begin{equation}
  H\left(\beta\right) = 
  \frac{2}{\pi\beta}\int\limits_0^\infty \mathrm{d}x~
  x \sin{\left(x\right)} \exp{\left(-\left(x/\beta\right)^{3/2}\right)} .
\end{equation}
An estimate of the mean ion density 
$\mathcal{N}_\mathrm{ion}$
is found by comparing this model with the data as follows.
The line shape of the measured EIT$_\mathrm{FP}$ signal $S(\Delta_{481})$ amounts to a 
convolution of the Holtsmark probability distribution 
with the EIT-signal shape $F_\mathrm{EIT}\left(\Delta_{481}\right)$
\begin{equation}
  S \left(\Delta\right) = 
  \int\limits_0^\infty\mathrm{d}E~\mathcal{P}\left(E\right) 
  F_\mathrm{EIT}\left(\Delta - \alpha E^2/2\right) . 
\end{equation}
In this integral, the EIT line 
(a $2\pi\times\SI{50}{\mega\hertz}$ wide Gaussian profile~\cite{fleischhauer2005electromagnetically}
as in our measurements),
is displaced by $-\alpha E^2/2$ according to the Stark-shift. 
Using the approximation~\cite{hummer1986rational} for the Holtsmark formula,
we find that the center of mass $C$ of the obtained line shape
linearly scales with the ion density as
\begin{equation}
  C = \SI{1.89e-9}{cm\volt^2} \alpha \mathcal{N}_\mathrm{ion} .
\end{equation}
This translates to a first estimate of the observed ion density of up to
$\mathcal{N}_{ion} \le \SI{1e10}{\centi\meter^{-3}}$ 
for an observed shift up to $2\pi\times\SI{250}{\mega\hertz}$ in the EIT$_\mathrm{FP}$ scheme.
Given the atomic ground state density 
and a Rydberg fraction of around \SI{2}{\percent},
this ion density matches \SI{27}{\percent} of the Rydberg density. 

We then approximate the ionization process with a rate equation.
Rydberg atoms with a number density $\mathcal{N}_\mathrm{Ryd}$ (assumed constant) 
collide with ground state atoms 
with a relative velocity $v$ and a cross-section $\sigma$,
contributing to the increase of the density of ions $\mathcal{N}_\mathrm{ions}$.
At the same time, ions leave the excitation volume with a rate of
$\gamma = \SI{1.5}{\mega\hertz}$, 
chosen to be the inverse of the transit time through the excitation volume 
\cite{demtroder2000linienbreiten}.
The corresponding rate equation is
\begin{equation}
  \dot{\mathcal{N}}_\mathrm{ions} = 
  \mathcal{N}_\mathrm{Ryd} \left(\mathcal{N}_{85}+\mathcal{N}_{87}\right) \sigma v - \gamma \mathcal{N}_\mathrm{ions} 
\end{equation}
and reaches its steady state at 
\begin{equation}
  \mathcal{N}_\mathrm{ions} = \mathcal{N}_\mathrm{Ryd} \frac{\left(\mathcal{N}_{85}+\mathcal{N}_{87}\right) \sigma v}{\gamma} .
\end{equation}
Substituting the ion density with the value from the first estimate
leads to an ionization cross-section of up to
$\sigma = \SI{1e-3}{\micro\meter^2} = 0.03 \cdot \sigma_\mathrm{geo}$,
where the geometric cross-section of the Rydberg atom is given by
$\sigma_\mathrm{geo} = \pi \left((n^\ast)^2 a_0 \right)^2$,
with an effective quantum number $n^\ast = n-\delta$ and quantum defect $\delta$.
Similar measurements in a pulsed experiment with an atomic beam showed 
$\sigma = 0.06 \sigma_\mathrm{geo}$~\cite{vitrant1982rydberg}, 
in very good agreement with our results.

\subsection{Cesium}
\begin{figure*}[htbp]
\centering
\includegraphics[width=\textwidth]{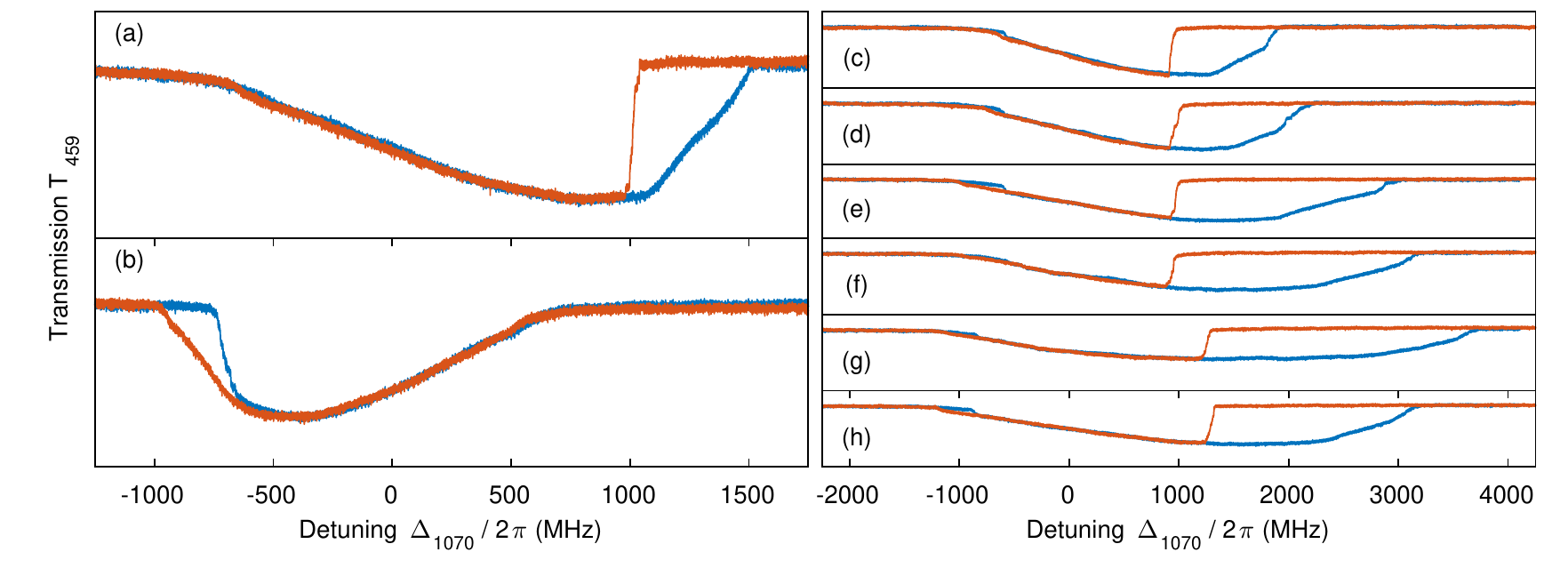}
\caption{
  {Cesium measurements.}
  Transmission spectra with frequency scans towards the color of the respective line. 
  \textbf{(a,b)}
  Comparison between (a) $23D$ and (b) $28S$ state. 
  \textbf{(c--h)}
  Effect on the bistability for the $23D$ state
  by RF modulation with a sine wave of 
  \SI{3.2}{\volt\per\centi\meter} amplitude and increasing frequency 
  (10, 20, 50, 100, 200 and \SI{500}{\mega\hertz}).
}
\label{fig:results:cs}
\end{figure*}

Further evidence
that the optical bistability is caused by electric fields is found
by analyzing the position of the bistability window
relative to the unperturbed resonance,
which gives the sign of the underlying interaction mechanism.
Figure~\ref{fig:results:cs}(a,b) shows our hysteresis spectra with optical bistability 
for the $23D$ and $28S$ states in $^{133}$Cs.
For both states,
the character of the Rydberg-Rydberg interactions 
is essentially repulsive~\cite{urvoy2015strongly, singer2005long}.
However,
the polarizability clearly changes its sign from 
$\alpha_\mathrm{23D} = \SI{-0.52}{\mega\hertz\per(\volt\per\centi\meter)\squared}$ 
to
$\alpha_\mathrm{28S} = \SI{+0.76}{\mega\hertz\per(\volt\per\centi\meter)\squared}$, 
as does the position of the bistable region.
We summarize these results as well as those from previous measurements in
Table~\ref{tab:polarizabilities}. 
When comparing the respective observation to the polarizability of each species and angular momentum states,
we find perfect agreement with our hypothesis.
\begin{table}[htbp]
\caption{\label{tabone}
Compiled signs of Rydberg Stark-shifts,
van-der-Waals interaction
and the actually observed position of the bistability.} 
\label{tab:polarizabilities}
\begin{tabular}{{l}{l}{l}{c}{c}{l}}
\hline \hline                            
Element & State & \multicolumn{1}{r}{vdW} & Stark-shift & Bistability & Source\cr 
\hline
$^{85}$Rb & $32S_{1/2}$ & $+$ & $-$ & $-$ & This work  \cr
$^{85}$Rb & $22S_{1/2}$ & $+$ & $-$ & $-$ & This work \footnote{Not shown.} \cr
$^{133}$Cs & $28S_{1/2}$ & $+$ & $-$ & $-$ & This work \cr
$^{133}$Cs & $23D_{1/2}$ & $-$ & $+$ & $+$ & This work \cr
$^{133}$Cs & $18--37P_{3/2}$ & $+$
\footnote{\textit{Essentially repulsive}~\cite{urvoy2015strongly, singer2005long}.} & $-$ & $-$ & \cite{carr2013nonequilibrium,de2016intrinsic}\cr
\hline \hline
\end{tabular}
\end{table}

Finally, we investigate
how the optical bistability is affected by external electric fields. 
Figures~\ref{fig:results:cs}(c) -- \ref{fig:results:cs}(h) show a significant increase in the width of the hysteresis 
as the frequency of the electric field is varied. 
The applied electric field amplitudes are small enough 
that the resulting Stark-shift
and the modulation frequency
are negligible with regard to the relevant energy scales of the Rydberg atoms.
The interparticle interactions of neutral participants are therefore not affected.
However,
charged particles are heavily influenced. 
In the range of frequencies used for the electric field modulation
(10 to \SI{500}{\mega\hertz}),
the free electrons from the ionization process are accelerated
and perform an oscillating trajectory, 
thus increasing the ionization rate by additional collisions. 
We believe that this explains why the width of the bistability region increases. 
Overall,
the susceptibility to electric field modulation contradicts a dipole-dipole interaction between Rydberg atoms,
and once more designates that ionization significantly contributes to the observed bistability.

Associated with a significant increase in the width of the hysteresis,
a small second hysteresis seems to appear on the opposite detuning sign,
observed only with an applied $E$-field
and above a certain threshold of frequency and amplitude combination.
We are not able to explain this observation at this point 
or to give more details about the underlying dynamics, 
and further theoretical and experimental studies are needed 
in order to fully grasp the microscopics of this complex system. 
It is worth noticing 
that a double hysteresis has recently been observed in a related experiment~\cite{ding2016non}. 

\section{Conclusion}
\label{sec:conclusion}
With the joint examination of two independent atomic species 
and combinations of principal and azimuthal quantum numbers
we are able to preclude Rydberg-Rydberg interactions 
as the mechanism 
responsible for the interaction shift in the phenomenon of optical bistability in thermal vapors.
At the same time,
we find strong evidence that electric fields produced by charges
originating from Rydberg ionization
are the dominant contribution 
to the observed effect. 
Our argumentation 
based on the results of two different experimental setups 
contradicts the previously suggested explanations published in
\cite{carr2013nonequilibrium,de2016intrinsic,ding2016non},
where the spectrum of experimental parameters overlaps with our settings.
 
In our first experiment,
we have applied two independent EIT schemes,
each addressing only one of the naturally abundant isotopes in a rubidium vapor cell.
The first scheme strongly drives the transition in $^{85}$Rb,
enabling the atoms to enter a bistable regime, while at the same time,
the effect on the $^{87}$Rb atoms is monitored by the second EIT scheme.
The choice of Rydberg states allows us to exclude dipole-dipole interactions.
We find both 
a deformation of the EIT$_\mathrm{OB}$ line 
and a shift of the EIT$_\mathrm{FP}$ line to the red.
These observations are in good agreement with the Stark effect caused by electric fields.
The fields originate from ionizing collisions of the Rydberg atoms,
and the estimated ion densities
and the required ionization cross sections are within a realistic range.

In the second experiment using $^{133}$Cs vapor,
we have shown that applying weak electric fields significantly alters
the width and position of the bistability.
Only charged particles being affected by such weak fields
and the distinct accordance of the sign of the polarizability for 
different states finally support our conclusion.

Further theoretical and experimental investigations are necessary 
in order to refine the microscopic model 
for the bistability observed in thermal vapor.
For example in Ref.~\cite{carr2013nonequilibrium},
a change in the fluorescence spectrum has been found,
which cannot be explained by Stark-shifted Rydberg states
or shifts due to (resonant) dipole-dipole interactions alone. 
This observation is attributed to superradiance by the authors.
The connection to bistability however needs further investigation
and additional processes have to be included in the model.
A more comprehensive understanding 
of the effects that shift the transition between the bistable states
will be beneficial for the progress towards future applications,
for example, ac electric field sensing~\cite{ding2016non,fan2015atom}.

\begin{acknowledgments}
The research leading to these results has received funding from the Carl-Zeiss Foundation,
the European Union's Seventh Framework Programme H2020-FETPROACT-2014 under Grant No.~640378 (RYSQ),
and BMBF within Q.Com-Q (Project No.~16KIS0129).
H.K. acknowledges personal support from the Carl-Zeiss foundation. We appreciate fruitful discussions with T.~Pfau and J.~P.~Shaffer.
Thanks to S.~Weber, C.~Tresp, and S.~Hofferberth for providing the code for the computations of pair states and polarizabilities
and to M.~Abdo and the Institute of Electrical and Optical Communications Engineering for support in developing the microwave circuitry.
\end{acknowledgments}

\bibliography{references}

\end{document}